# Deployment of an Innovative Resource Choice Method for Process Planning

A. Candlot[1], N. Perry[1], A. Bernard[1] (1), S. Ammar Khodja[1]
[1]Institut de Recherche en Communication et Cybernétique de Nantes
Ecole Centrale de Nantes, Nantes, France

**Abstract**
Designers, process planners and manufacturers naturally consider different concepts for a same object. The stiffness of production means and the design specification requirements mark out process planners as responsible of the coherent integration of all constraints.
First, this paper details an innovative solution of resource choice, applied for aircraft manufacturing parts. In a second part, key concepts are instanced for the considered industrial domain. Finally, a digital mock up validates the solution viability and demonstrates the possibility of an in-process knowledge capitalisation and use. Formalising the link between Design and Manufacturing allows to hope enhancements of simultaneous Product / Process developments.

## 1 INTRODUCTION

The influence of product definition tends to be accurately reflected in the whole enterprise. Its life cycle implies new monitoring tools for cost management [1], for planning or for process definition and deployment. Among the whole product life cycle, this paper focuses on a particular sensitive point of manufacturing: process planning.

Between design and manufacturing, the process planner receives from one side the definition of the geometry and from the other the process capabilities of the workshop. It is illustrated on Figure 1. The efficiency of this activity directly influences the possible flexibility to configure the process according to the product definition.

Thus, the worth of a simultaneous definition of products and processes directly depends on the size of the expected production. When the product batch is important, more time can be consumed on a process optimisation during process planning. But if the product batch is too small, the customization of the process cannot be economically provided and the constraints absorption has to be assumed by the process-planning phase [2] that could then require an optimisation.

In case of small batches, design geometry barely gives production requirements else than nominal dimension. On the other side the databases coming from the production hardly reflect the real experimental feed back from previous work pieces on the capability to manufacture specific recurrent types of geometry. These difficulties of the process plan can be distributed on three phases.

- The first of them analyses the manufacturability of parts and transforms by computation the design geometry in a mathematically equivalent semantically different manufacturing geometry.
- The second phase concerns the construction of the setup structure. Faces and associated processes are distributed to setups. The existence of capable process must be ensured for each manufacturing geometry element [3].
- The last phase consists in an optimisation of parameters respecting the constraints elaborated in the two previous phases and calculates optimal cutting conditions and tool trajectories.

Nevertheless, process planners naturally work simultaneously on these three aspects. Decisions are closely tied and it is a major difficulty to organise them in a robust general sequence [4] [5].

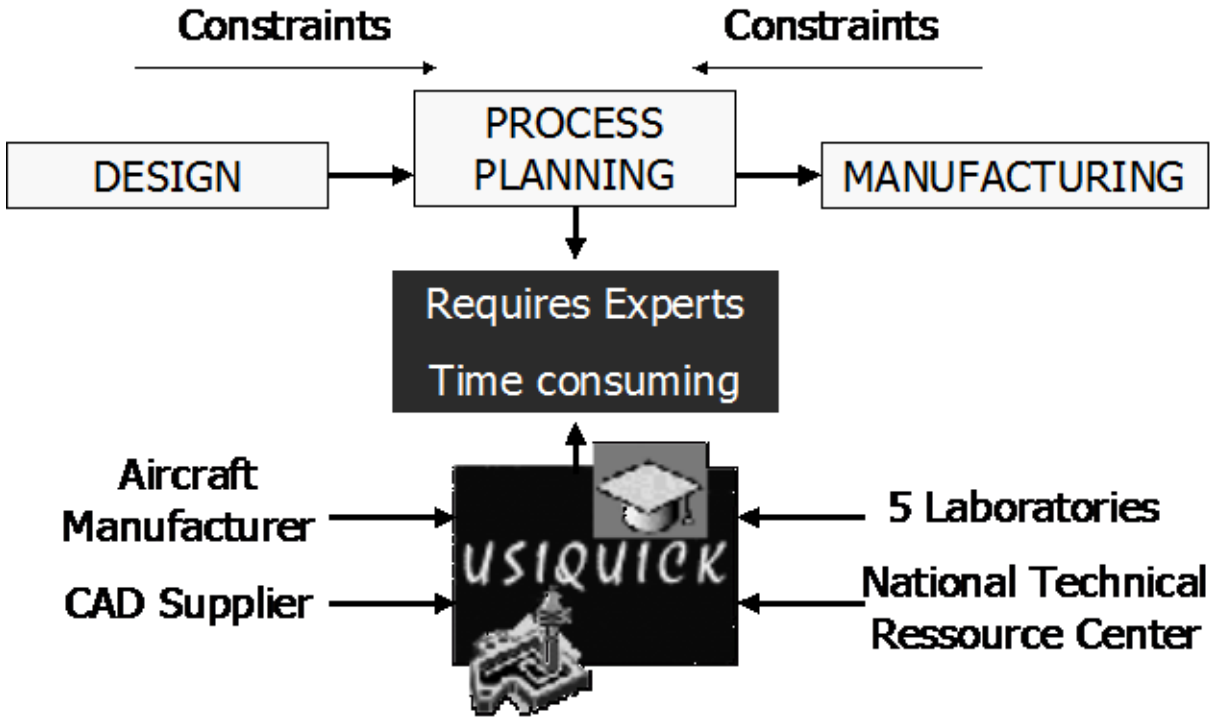

Figure 1: Role of process planning in the PLC [6] and interest of the USIQUICK project [7]

The association in a same database of geometrical feature definitions, tools and acceptable domains of cutting conditions would allow process planners to choose and tune efficiently the deployed resources. Allowing automation of recurrent steps, it should lighten the work of the expert and ensure at long term a better coherency between products and processes. A theoretical method called OSE (French acronym for Tool, Sequence, Feature) has been proposed by [11]. The aim of the paper is to assess its deployment possibility in a CAM tool for small-batch aircraft manufacturing.

At first the scope of the project is detailed. Then in a second part, theoretical concepts are introduced and defined. Adaptations and instances concerning the reality of the project context are detailed. The following section details the general requirements induced by the industrial partners, for the process plan activities in one side and to comply with the already existing software solution. Finally the deployment of the first OSE list is detailed and analysed.

## 2 PROJECT CONTEXT

### 2.1 Partners and scope

The works presented here are part of the output from a project financed by the French Ministry of Industry, called USIQUICK [7]. It involves eight partners:

- An aircraft manufacturer specifies the expected results and proposes its expertise on process planning.
- A CAD/CAM development leader plans the industrialisation of outputs in its software solution.
- Five laboratories ensure the scientific coherence of the project and propose innovative solutions to solve strategic locks.
- A French-government institute analyses the possible use in other fields and proposes extra test cases and tool databases.

The project focuses on the definition of milling process plans in aircraft manufacturing with a high amount of re-engineering. It implies particular geometries and processes. Because of the sizes of batches induced by frequent re-engineering, this activity must be fast and flexible. Solutions must be almost but not necessarily completely optimum. This particularisation of the problem made compete the theoretical solutions and the integration reality.

The difficulty is then to identify what are the knowledge elements that have to be kept customisable and that have to be definitely validated and integrated. Concepts have to be firstly identified and extracted, secondly structured and formalised and then refined to an accurate level of maturity[8] [9].

### 2.2 Knowledge-based Engineering Tool

In order to optimise the information flow from design to production, a three-step method is proposed [7] according to the three phases highlighted in the introduction.

- Transformation phase: an analysis of the part to compute a maximum of information registered in an appropriate level of feature. In this phase computer assess the machinability of faces by evaluating OSE parameters.
- Preparation phase: the synthesis templates of the previous phase are presented to the user. Then with appropriate tools, the process plan skeleton can be built and constrained. The similarities between OSE parameters help to group faces according to their accessibility in order to constitute setups.
- Automation phase: the unconstrained choices are automatically optimised and a complete documentation is processed by the system.

These phases would become the three major elements of an engineering tool based on the formalisation of expert knowledge.

In product life cycles, engineers face increasing information flows that are difficult to handle for decision-making. These flows come from different experts or departments or from previous works. The control of these flows is called knowledge management and may require supporting software. Tools designed in this context are called Knowledge-based engineering (KBE) tools [10].

## 3 DEFINITIONS

### 3.1 General Semantic

The OSE model links three databases in one: the cutting sets, the machining conditions and the admissible geometry conditions as illustrated in Figure 2. Each OSE is a compatible combination of an element of each table.

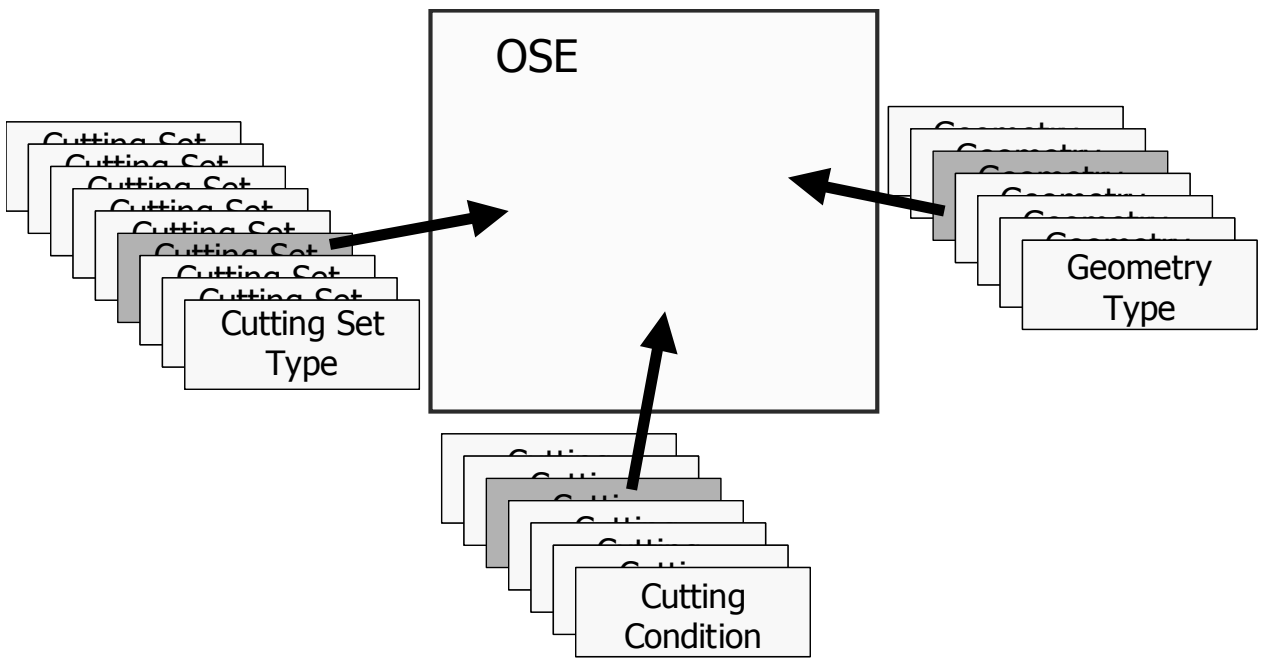

Figure 2: OSE Design Principle

In order to clarify the terminology, the initial thesis from [11] was dealing with tools, sequences and entity or feature but these concepts had to be adapted to the industrial context.

The aircraft-manufacturing partner prefers to manage tools with their attachment to the machine. Tools thus become cutting sets including fasteners to ease the tool-gauge dimension management in the workshop.

Possibilities concerning feature recognition did not allow managing high-semantic element like notably pockets [12]. Computation is easier with geometrical elements. The project obtained a list of manufacturing geometries, corresponding to the machining possibilities of milling tools and machines.

The definition of a coherent list of geometrical elements is one of the main issues of the transformation phase [13]. An analysis of process possibilities gives a list of typical machinable geometries.

It has been obtained by the analysis of a set of typical parts. [14] propose a scientific approach of this typology analysis applied on die manufacturing that is used to refine the USIQUICK project list. This list of types that has been selected to be deployed on the 5-axis aircraft-manufacturing project is the following:

- Plan
- Cylinder
- Cone-Shaped Surface
- Ruled Surface
- Constant-Radius Sweeping Surface
- Unspecified

These features must be considered exclusive even if they classically are not. For example, a plan must not be considered as a ruled surface and a ruled surface is not a plan. A given geometry has to belong to the more restrictive group it can.

These categories have been sorted out on three real case parts. The results are presented in the Table 1. It actually points out that these categories could be further refined. Notably some works in the project aim to differentiate real cylinders constrained by design from manufacturing fillet that should be suppressed to become an edge attribute. Unspecified features still represent a major number. They are usually managed by sweeping (with a special case of constant radius sweeping that has been separated). Works are analysing how to transform some of the unspecified features in ruled faces in case of minor differences. It could result in a new category or increase the number of ruled faces.

| | Part 1 | | Part 4 | | Part 7 | | TOTAL | |
|---|---|---|---|---|---|---|---|---|
| Plan | 50 | 14.71 | 66 | 29.46 | 53 | 11.06 | 169 | 16.20 |
| Cylinder | 109 | 32.06 | 73 | 32.59 | 76 | 15.87 | 258 | 24.74 |
| Cone-Shaped Surface | 15 | 4.41 | 0 | 0.00 | 14 | 2.92 | 29 | 2.78 |
| Ruled Surface | 13 | 3.82 | 25 | 11.16 | 38 | 7.93 | 76 | 7.29 |
| Const.-R Sweeping Surf | 9 | 2.65 | 21 | 9.38 | 88 | 18.37 | 118 | 11.31 |
| Unspecified | 144 | 42.35 | 39 | 17.41 | 210 | 43.84 | 393 | 37.68 |
| TOTAL | 340 | 100.00 | 224 | 100.00 | 479 | 100.00 | 1043 | 100.00 |

Table 1: Number of faces identified by type on three examples

These geometrical elements simplify the definition of sequences that in the original work of [11] should have been lists of operations. As faces are considered, the sequence is closer to an operation and thus, only a single cutting-condition set has to be considered. A cutting-condition set and associated parameters of application are here called extended cutting conditions.

The construction of wider sequences is managed in a later step of the preparation phase when similar operations of same setups are grouped. After the breakdown of phases in setups, each setup contains faces and associated processes that can be compared to identify similarities in potential OSE. An optimisation algorithm can then regroup complementary faces. For example, a pocket flank of several faces can be recognised if all the faces have a same candidate in its OSE list.

This late recognition is the consequence of the collaboration of two working systems. The human system could recognise high semantic elements in the early steps and then breaks it according to smaller machining operation. It is a top / down semantic approach [12]. The computer system does not have access to meaning and cannot instinctively link particular typologies with conceptual process principles. It can only manage logical information as geometry and compare parameters. If it has access to relevant parameters, it can rebuild higher-level structure. It is a bottom / up syntax-based approach.

In a nutshell, the role of OSE is to help the expert to formalise its process planning knowledge in computer-understandable pieces [15]. Then this embedded information is sent in the early computer phases to help its parameter recognition for a maximum use of computation. The information flows selected are:

- USIQUICK features, or geometry sets, defined by their capacity to be managed by both human and computers and according to a significant sample of parts
- Extended cutting conditions making the link between manufacturing geometry parameters and process capabilities.
- Cutting set types, grouping tools according to their actual attachment and to their main geometrical and manufacturing characteristics. The final selection depends on the cutting conditions required.

### 3.2 Main Association Principles

The introduction of types avoid combinatory explosion. For example, two plans defined with a different topology but corresponding to a same kind of process can then be managed in a homogeneous way if they are identified only by their common parameters. Planar faces can be for example sorted out according to their outlines and their manufacturing mode. Thus all plan unlimited by other faces and requiring only a roughing operation could be linked with general surfacing conditions. Other plans limited by other faces and requiring a finishing operation could be associated with the common process used for pocket fonts.

Thus, for one machining condition, two lists of checks group geometry elements in one hand and cutting set in the other. Possible couples of cutting sets and geometry are obtained.

For a specific work part, the checks are ran to identify which faces are belonging to each geometry set or "family". A list of OSE containing these sets is then obtained and with them, relevant potential cutting sets. The whole process is detailed in the Figure 3.

The main difficulty relies in the accuracy of the set definitions. The first expert populating the database must ensure that each rule set is coherent with the whole database as for instance in the Figure 4. This already difficult task becomes even more difficult if several experts have to work together on a long period. The first reflex is to try to fix the system in a definitive and robust configuration. But in this case, the OSE system cannot ensure to reflect the enterprise process capability. By definition, the KBE tool must stay open to process-planner tuning.

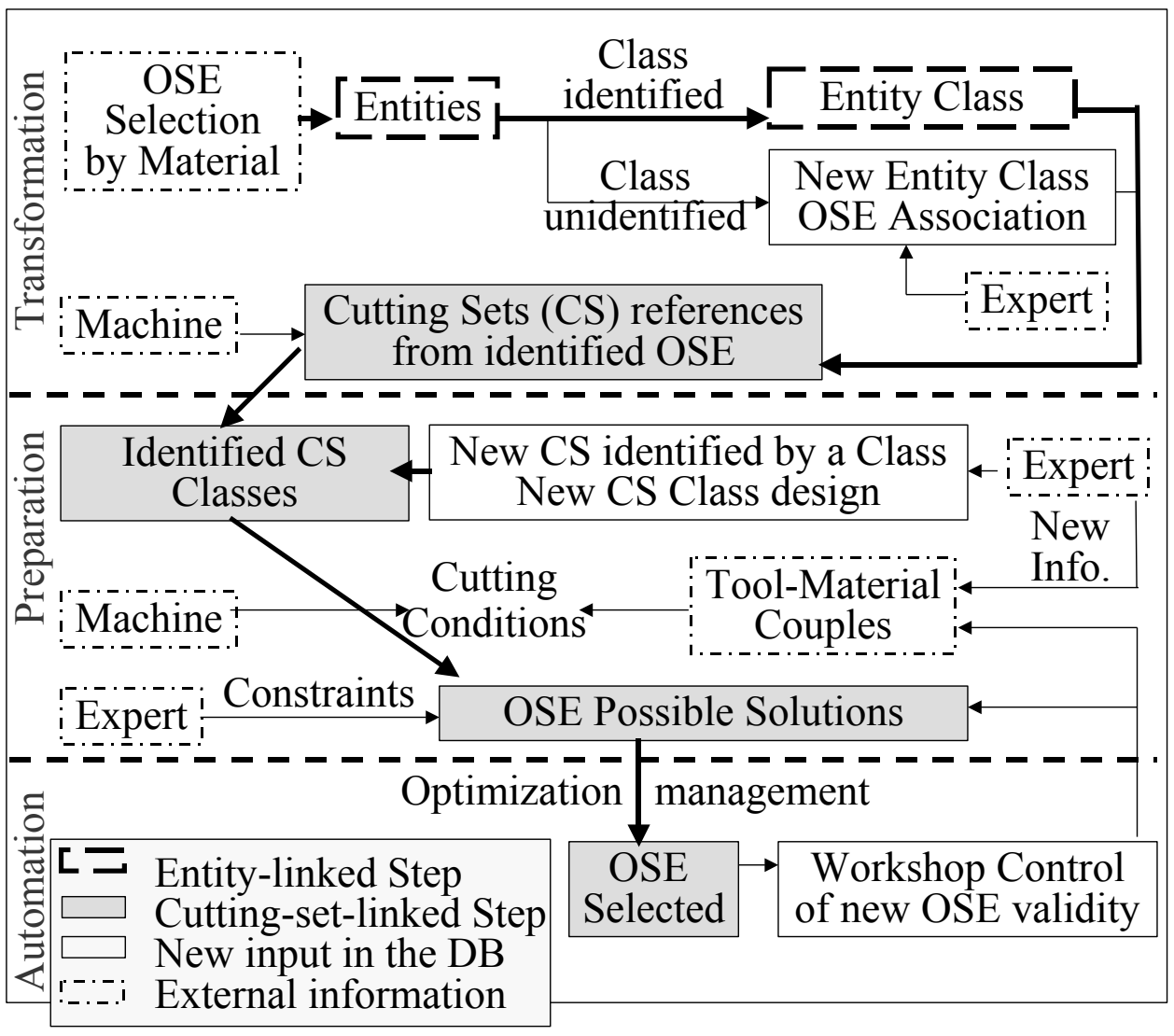

Figure 3: OSE used in USIQUICK process

Thus a method is required to support experts in the refinement of the OSE database during its construction and later during its use. The informal phase of the MOKA method has been selected to support the deployment of the project. The resulting knowledge base could be used later to sustain the system during its life cycle.

This method and the general expert requirements for the final KBE tool it contributed to organised are detailed in the following section.

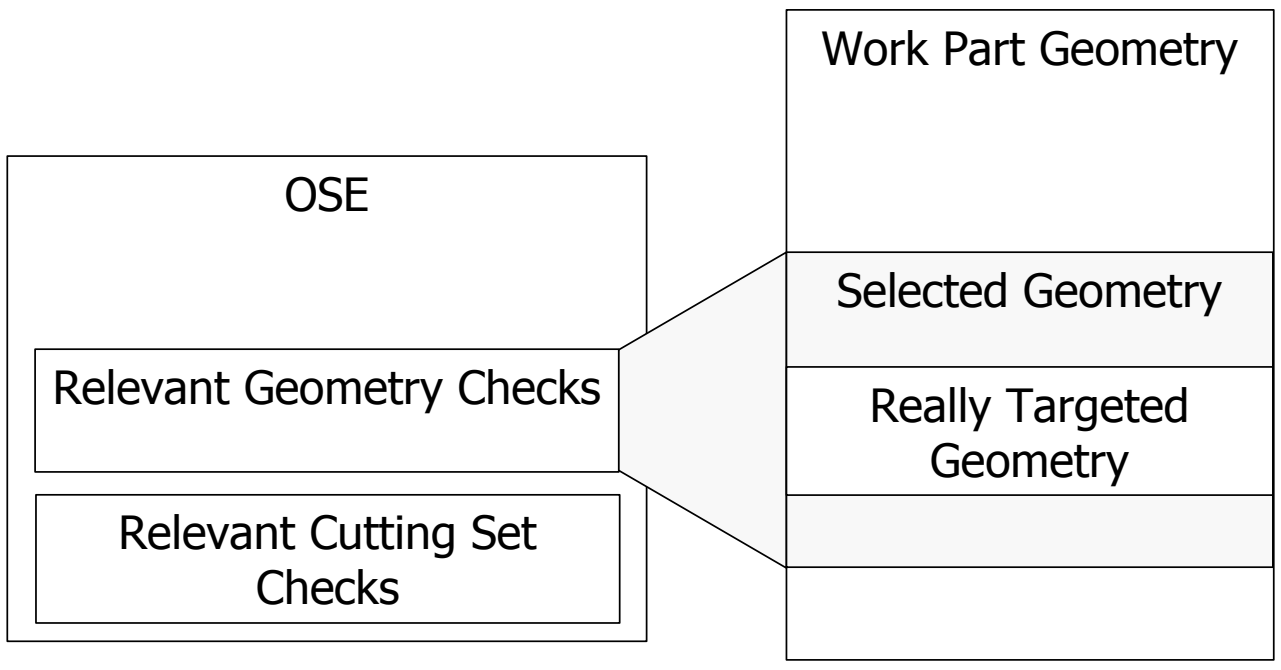

Figure 4: Difficulty to assess rule relevance

## 4 ANALYSIS OF DOMAIN AND SOFTWARE CONSTRAINTS

### 4.1 Methodology and tools for knowledge extraction, structuration and formalisation

The relevance of the database depends on its coherency towards industrial requirements and the actual possibilities of software solution. The mapping of these two models is difficult to obtain from the available resources mainly expressed in natural language. This lack of formalisation induces an increased difficulty to justify a well-formed database design.

To face this issue, two complementary works of formalisation have been run. The first analysed the specifications to construct a consensual relevant data model and the activity flow to handle it. The formalism chosen was class, activity and sequence diagrams from UML 2 standards [16]. The second work aimed to list the key instances of the data model in the text resources in order to first validate the previous consensual model and then to build a knowledge base that would pilot a coherent deployment of the rule and database. The informal phase of the MOKA methodology assumes this task [17].

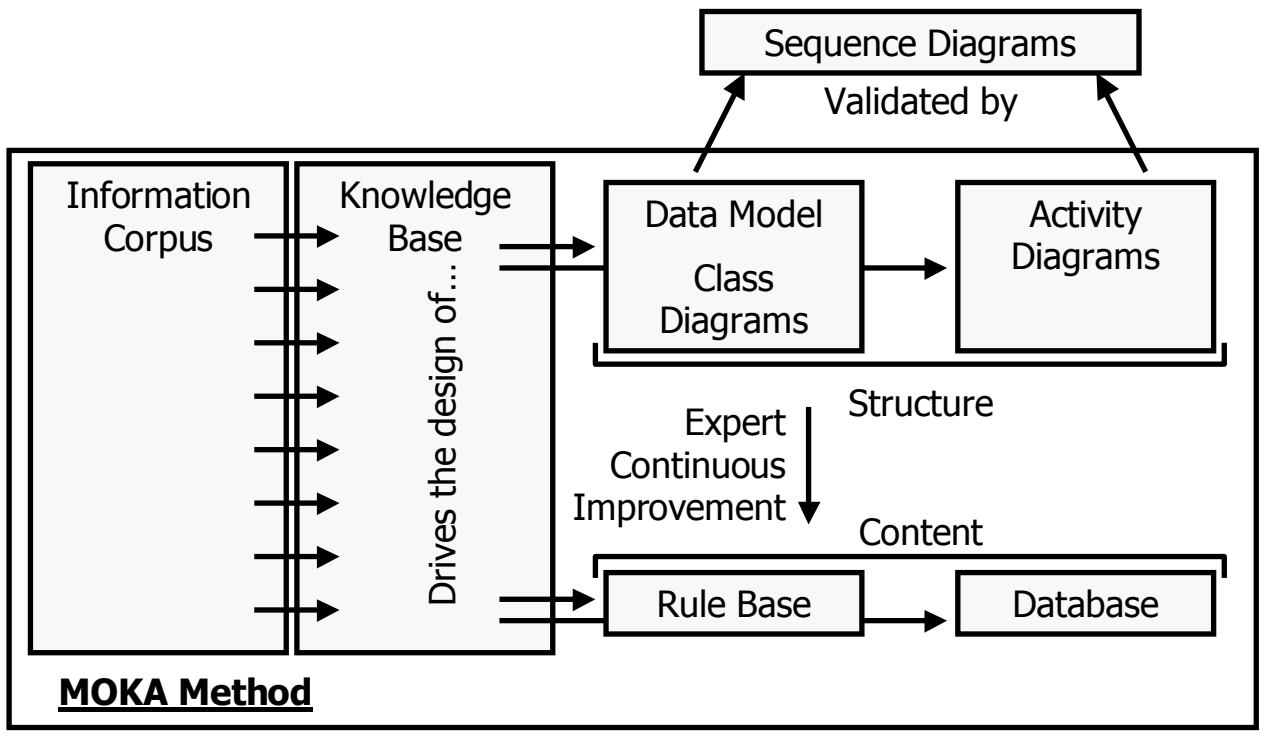

Figure 5: Synthesis of the knowledge management for creating a knowledge-based engineering tool

This methodology is composed of four main phases that are the project definition, the product representation, the process representation and the generation of the knowledge base. The extraction tool used is Pc-Pack [18].

The product model is divided in functional and structural breakdowns. These two diagrams have to be bound. Then identified constraints are allocated. The process model is made of an activity breakdown to which rules are allocated. Finally the knowledge base groups the product and process models and is completed by the links between them. For each element of all these representations, an ICARE card is produced to capitalise relative knowledge.

Figure 5 sums up the knowledge groups handled by these two modelling activities. These elements constitute the skeleton of the final KBE Tool. The “structure” represents the tool itself and the “content” represents the information extracted from expert documents or expert interviews and that will populate the containers of the structure.

### 4.2 Industrial Domain Requirements

The main expectation of the industrial partner was to separate usual decisions of special-case ones. The computer would have to manage the maximum of the firsts and concentrate the need of external decisions in specific highly expert points.

The consequences of this choice require:

- To maintain alternatives until an expert validation.
- To take decision only when required information is available [19].

The formalised expert knowledge has to be introduced as early as possible to reduce possible combinatory explosion. Thus OSE should be used as proposed at the early steps of the feature construction, during the transformation phase. It participates to the machinability validation.

The second added value of OSE is to help to capitalise and manage alternatives during the preparation phase. The chosen implemented formalism must be carefully designed to help the expert understanding. The semantic of the three information flows represented in each OSE must be clear to understand.

### 4.3 Software Infrastructure

Two points guide the influence of the project on the software solution infrastructure:

- Try to manage only one alternative of an object.

For confidentiality reasons

- Limit code modification and use or inherit from existing objects.

Discussion on the modification of the data model, and solutions are not presented for confidentiality reasons.

## 5 OSE DATABASE MODELING PROCESS

### 5.1 Initial State

To construct a first OSE Database for the project, the process plan of a simple part (24 faces, see Figure 6) is studied. It starts by the identification of the geometry type for each face. Then a list of the choices that led to the operations concerning a particular face is constituted. It gives a base for a first OSE candidate.

The next step is to formulate the choice rules with project identified keywords. It contributes to highlight the relevance of what has been identified in the knowledge or database.

It has been also decided to separate rules in “if… then… else…” in two rules “if… then… “. In this case an OSE is separated in two. There are two reasons to this decision. First, OSE aims to capitalise the favourite expertise of an enterprise. Exceptional case should stay out of the system and the process planner should always keep the possibility to adapt results according to his interpretation of the context. So there is no need to store all particular cases. The second reason resides in the formalism choice. To handle easily parameters, checks have been selected. A check (that can be represented by “if … then ok / not ok”) constitutes an elementary brick of knowledge and can be applied to different OSE with less modifications than a more dedicated rule in “if… then… else…”.

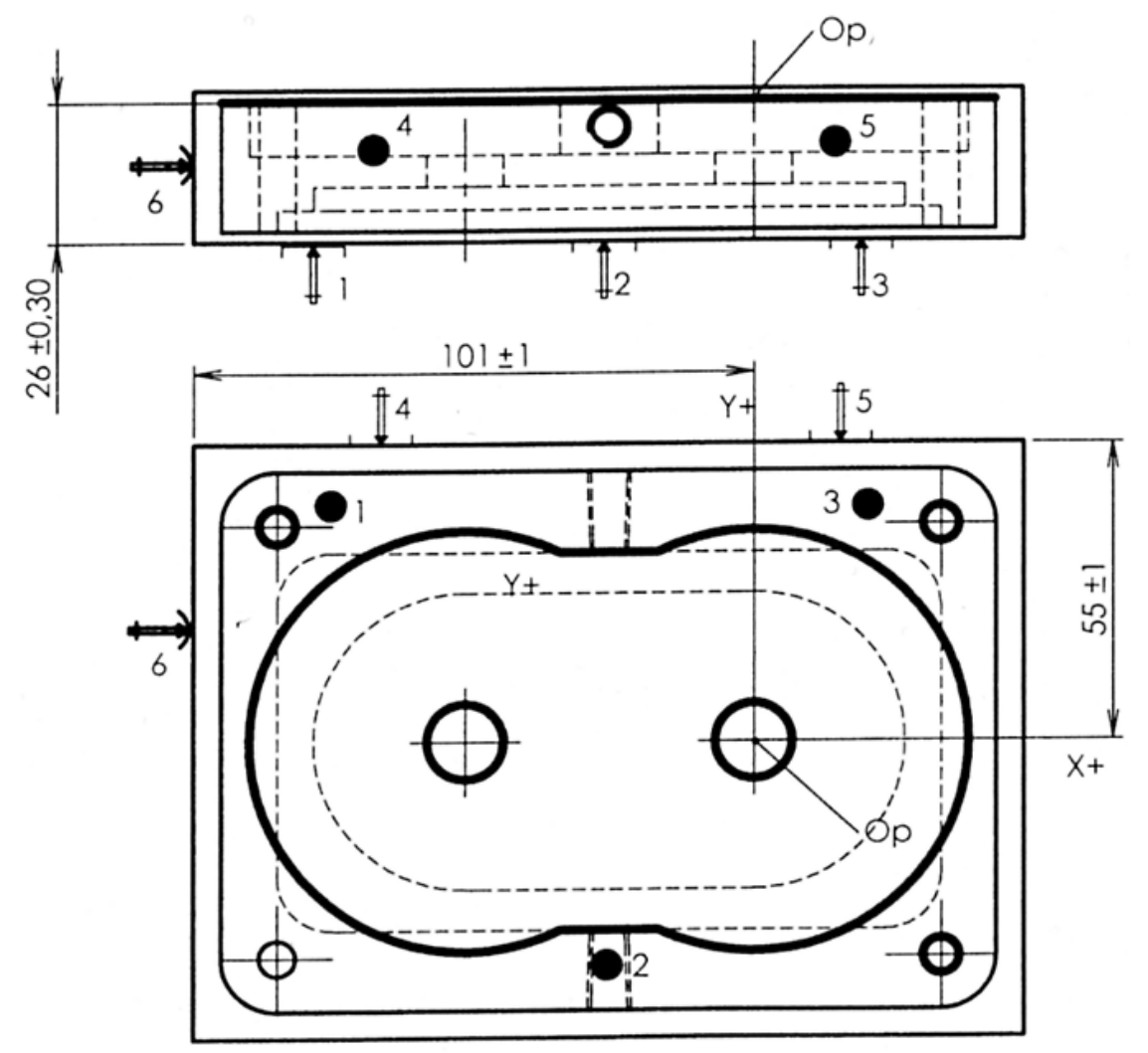

Figure 6: Example Pump carter part

The first list of OSE is then completed by a "What if" analysis [20] on each parameter choice in order to identify with manufacturing experts what are the combinations that are relevant and need to be capitalised. For example, if an OSE describes how to process a specific plan with surfacing parameters in roughing, the database builder may analyse what would be the situation if finishing were required.

Finally, the OSE database obtained can be compared, completed and validated by the rule and database of previously described MOKA analysis.

The number of parameters rapidly induces a high number of OSE. To lighten this number, the priority is given to the link between geometry and compatible tools. In a second time, for a given couple, the available manufacturing parameters are indicated. It allows a flexible user mode in three levels:

- At first and by default, the best extended-cutting-condition configuration is selected.
- If the software or the expert detects a problem, the expert can access all the other configurations that have been validated (by experts or by experiment feedback) and can select one.
- If none of the proposed solutions is satisfying, the expert can tailor his own solution.

### 5.2 Complements on Parameters

The configurations of manufacturing parameters correspond to a selection in the parameter breakdown of "extended cutting conditions" of the modified OSE previously introduced. There are three categories:

- Manufacturing types (end manufacturing, flank manufacturing, sweeping, drilling) and modes (roughing, semi-finishing, finishing);
- Trajectory strategy [21] (Forth, Back & Forth, In Out Spiral, Out In Spiral, Normal Drilling, Deburring, Flank, Sweeping…); some of these alternatives are not refined enough to differentiate efficiently faces more than the manufacturing types. This category is used in the automation phase of the project that has not been yet completed.
- Tool / Material couple - TMC (Cut Material, Cutting Material, Cutting Conditions Constraints, Lubrication)

The cutting set types is defined through the following list of principle parameters (can be refined or extended according to experts):

- Dimension ranges: Tool Diameter, Cutting Length, Tool Length, Tool End Radius.
- Cutting conditions ranges: Cutting Speed (global, by tooth), Advance ( X & Z), Feed Rate
- Cutting Material

And finally, the following parameters must be evaluated for each face:

- Outline openness; if edges are "open" (concerning less than 180° of material) or "closed" (concerning less than 180° of material) influences the accessibility of the face.
- Accessibility directions (Single Vector, two opposite Vectors, N. Vectors (continuously)); it is used to evaluate the potential manufacturing type and to build the setup breakdown. A single face may have several directions. It must be also specified if a direction is compulsory or not.
- Accessibility dimensions (End Accessibility: to indicate the smaller dimension of the face; Flank Accessibility: to indicate the longer dimension of the face; Global Accessibility: to indicate the depth level of the face in the part; Fillet Problem or Minimum Curve: that constraints tool end radius); it selects possible cutting set types and tools for each type. A dimension box can also be estimated to check general interferences with cutting set envelope.
- Potential Manufacturing Type; it is deduced from accessibility results.

Some other specific parameters could be added for each geometry type. They are not used in the first experiments.

The organisation and combination of these parameters are presented in a first-draft template on Table 2. Figure 7 indicates the semantic of zones.

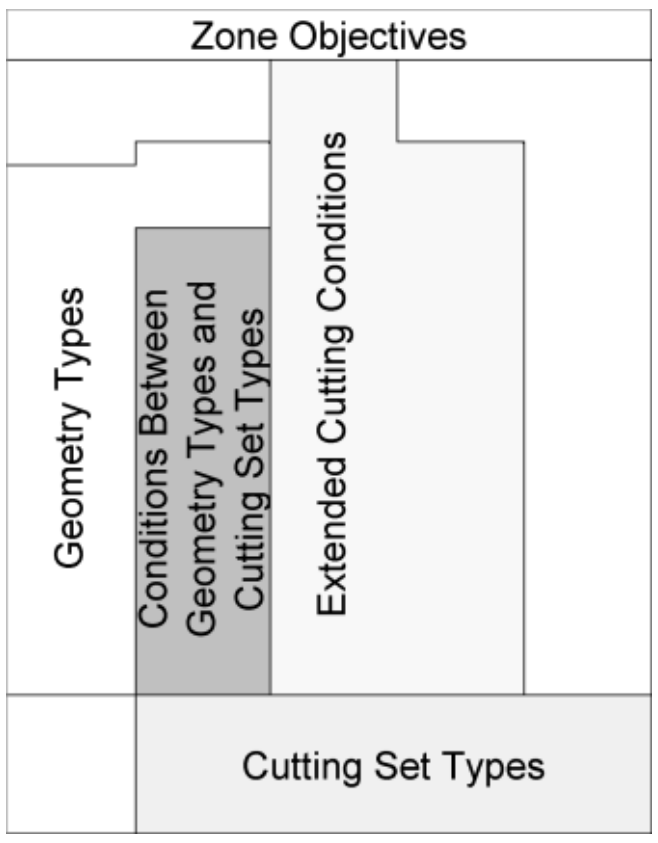

Figure 7: OSE Zones of the Table 2

The middle zone "Conditions Between Geometry Types and Cutting Set Types" is used to drive and ease the deployment of checks in the software solution. The checks contained by the OSE and applied to a tool database can be illustrated as follow:

Tool Diameter < End Accessibility (1)

If the condition is checked for a cutting set, then this element is a valid candidate for the OSE. The complete rule is of the form:

If (Tool Diameter < End Accessibility) then (Cutting Set is applied) (2)

So if the OSE called Feat 1 in the Table 2 has manufacturing type and mode valued at "end manufacturing" and "roughing" and is part of a work piece compatible with tool / materiel couples 1 and 3, it could be written as follow:

If (Face. Mfg Type = End Manufacturing and

Face. Access = Single Vector and

Face. Access Type = Compulsory)

Then (Feat 1 is compatible with Face) (3)

If (Tool Diameter < End Accessibility and

Tool Length > Global Accessibility and

Minimum Fillet Radius >= Tool End radius)

Then (Cutting Set verifies geometrical compliance) (4)

If (Tool Mfg Type = End Mfg and

Tool Mfg Mode = Roughing and

Tool TMC = TMC 1 OR TMC 3)

Then (Cutting Set is a candidate and

Cutting Condition Calculus Priority = Qmax) (5)

The first formula corresponds to the geometry-type definition. The two last formula highlight the two kinds of attributes supported by the geometry that are geometrical attributes and manufacturing attributes and corresponding to the cutting set and extended cutting condition choices. Remembering previous sections, these attribute values are obtaining during the transformation phase. The equality in formula (5) must be understood as follow: “the list corresponding to this attribute of the cutting set must contain the following elements”. It allows defining cutting sets compatible with several modes or types. For example, a same generalist cutting set could be used for roughing and finishing. The OSE can also be refined with trajectory types either at this phase or later in the process planning details definition.

### 5.3 Normal Use

The OSE database is difficult to tune and definitely requires synthesis tables. Once deployed, after a ramp up period, the system should be stable for a given kind of manufacturing parts. New tools should naturally be sorted out by the already existing OSE. Modifications of checks using existing parameters should be easy, even if the global coherence of the system would be hard to assess. For example, a modified OSE could replace another one for an unchecked configuration.

Moreover, users may need to refine categories by creating new ones (precise for example two kinds of cylinders: concave and convex or fillet and normal). These modifications that should be easy for cutting sets (their parameters are accessible by tool database) could be much more difficult for geometry types or extended cutting conditions. It would require new algorithms and so an extra programming effort. It highlights the importance of mock-up activities.

## 6 CONCLUSION

Compared to classical tool databases, OSE benefits of a link between geometry and processes. This link offers several advantages:

- From manufacturing point of view, it gives the opportunity to manage cutting sets or tool databases according to real process needs and to tune their number to the minimum. Tool demand is directly driven by the OSE uses.
- From process planning point of view, which is the first aim of the system, it allows capitalising expert knowledge. Actually, process planning is so complicated that the “style” of expert can be recognised from a part to another. If the OSE database is sufficiently documented and justified, it will contribute to rationalise this activities and help to transfer knowledge from a person to others.
- From design point of view, it could help to check the relevance of a design toward the process possibilities. Designers could use the tool function corresponding to the automatic transformation phase to analyse if the studied part is well covered by OSE.
- From computational point of view, the OSE database introduced here is designed to maximise the tasks that a computer can handle. This knowledge breakdown offers expert analysis to the early automatic steps.

The main drawback outlined in the previous sections is the amount of work between the specifications of the KBE tool and the corresponding OSE database and the real deployment in an end user environment. If batch algorithms validate scientific and industrial hypothesis, there is still an important factor determining the viability of the system to assess: the human acceptance. For this reason, ergonomics must be carefully studied:

- The expert wants to control the decision of software. There is a confidence link to build that can go through an interactive documentation. This automatic-decision control can become a learning tool.
- The different reviews that can be performed during the process plan and notably the final validation require an overview of a studied case.
- Database modifications are difficult to estimate. An overview of the OSE database could be useful. The Table 2 is a first draft of such a solution.

Between the competitive study and the industrialisation phase, the main concepts have been validated and the associated typologies are in the late refining phases. For instance, the level of geometry complexity for features is fixed but the selected elements that will be managed can still evolved. The future works will aim at organising this object for a maximum user understanding.

## 7 ACKNOWLEDGMENTS

The different partners of the USIQUICK project are greatly acknowledged for the existing synergy developed in this project. The CRAN Laboratory team is specially thanked for their transformation phase mock up.

The two industrial partners are also thanked.

| Geometry Checks Identify the type of the geometry analysed | | | | | | | | Tool % Geom Checks Compare Tool arguments with Feature geometrical arguments | | | | | Tool % MfgArg Checks Compare Tool arguments with Feature Manufacturing arguments | | | | | | | | | | | | | | | | | | | | Inde-pendant Filter | | | |
|---|---|---|---|---|---|---|---|---|---|---|---|---|---|---|---|---|---|---|---|---|---|---|---|---|---|---|---|---|---|---|---|---|---|---|---|---|
| | | | | | | | | Conditions (> ; < ; <= ; >=) | | | | | End Mfg | | | Flank Mfg | | | Sweeping | | | Drilling | | | Non Critic for Mfgability | | | | | | | | Define Vc | | | |
| | | | | | | | | | | | | | Roughing | 1/2 Finishing | Finishing | Roughing | 1/2 Finishing | Finishing | Roughing | 1/2 Finishing | Finishing | Roughing | 1/2 Finishing | Finishing | | | | | | | | | | | | |
| **Feature** | | Comment | Single Vector | 2 opposite | N Cont. Vect. | COMPULSOR | POSSIBLE | End Accessibility | Flank Accessibility | Global Accessibility | Fillet Pb / Min Curve | Including Box / Dimension | RoEMfg | sFEMfg | FEMfg | RoFMfg | sFFMfg | FFMfg | RoSw | sFSw | FSw | RoDr | sFDr | FDr | Forth | Back & Forth | In Out Spiral | Out In Spiral | Normal Drilling | Deburring | Flank | Sweeping | TMC 1 | TMC 2 | TMC 3 | TMC 4 |
| Feat 1 | Plan | Strict. End Mfg | x | | | x | | Tool Diam | | Tool Length | Tool End Rad | | x | x | x | | | | | | | | | | x | x | | | | | | | | | | |
| Feat 2 | Plan | End or Flank Mfg | x | | x | | x | Tool Diam | Cutting Length | Tool Length | | | x | x | x | x | x | x | | | | | | | x | x | | | | | x | | | | | |
| Feat 3 | Plan | Strict. Flank Mfg | | | x | | x | | Cutting Length | Tool Length | Tool Diam | | | | | x | x | x | | | | | | | | | | | | | x | | | | | |
| Feat 4 | Cylinder | Double Access | | x | | | x | Tool Diam | Cutting Length | Tool Length | | | x | x | x | | | | | | | x | x | x | | | | | | x | | | | | | |
| Feat 5 | Cylinder | Single Access | x | | | x | | Tool Diam | Cutting Length | Tool Length | | | x | x | x | | | | | | | x | x | x | | | | | | x | | | | | | |
| Feat 6 | Cone-Shaped Surface | | | | x | x | | | Cutting Length | Tool Length | Tool Diam | | | | | x | x | x | x | x | x | | | | | | | | | | x | x | | | | |
| Feat 7 | Ruled Surface | Double Access | | x | | | x | | Cutting Length | Tool Length | Tool Diam | | | | | x | x | x | x | x | x | | | | | | | | | | x | x | | | | |
| Feat 8 | Ruled Surface | Single Access | | | x | x | x | | Cutting Length | Tool Length | Tool Diam | | | | | x | x | x | x | x | x | | | | | | | | | | x | x | | | | |
| Feat 9 | Ruled Surface | General | | | x | | x | | Cutting Length | Tool Length | Tool Diam | | | | | x | x | x | x | x | x | | | | | | | | | | x | x | | | | |
| Feat 10 | Const.-R. Sweeping Surface | | | | x | x | | | | Tool Length | Tool End Rad | | | | | | | | x | x | x | | | | | | | | | | | x | | | | |
| Feat 11 | Unspec. | Particular Cases | | | | | | | | | | | | | | | | | | | | | | | | | | | | | | x | | | | |

( A Line corresponds to an OSE)

| Tool Diam | Cutting Length | Tool Length | Tool End Rad | | Qmax | Vc max | Vc max / Ra / a mini | Qmax | Vc max | Vc max / Ra / f mini | Qmax | Vc max | Vc max / Ra / a or f mini | Qmax | Vc max | Vc max / Ra / ae mini | Cutting Material | Vc | fz | f | a | Q | Cutting Sets | Fastener Type | Max Diameter | Min Diameter | Teeth Number | Lubrif. |
|---|---|---|---|---|---|---|---|---|---|---|---|---|---|---|---|---|---|---|---|---|---|---|---|---|---|---|---|---|
| | | | | | | | | | | | | | | | | | | | | | | | Type 1 | | | | | |
| | | | | | | | | | | | | | | | | | | | | | | | Type 2 | | | | | |
| | | | | | | | | | | | | | | | | | | | | | | | Type 3 | | | | | |
| | | | | | | | | | | | | | | | | | | | | | | | Type 4 | | | | | |
| | | | | | | | | | | | | | | | | | | | | | | | Type 5 | | | | | |

Table 2: First Draft of OSE Database for USIQUICK Project